\documentclass[journal]{IEEEtran}
\usepackage{multirow}

\ifCLASSINFOpdf
  \usepackage[pdftex]{graphicx}  
\else
\fi

\begin{document}

\title{Identification of Phase-Locked Loop System From Its Experimental Time Series}

\author{Mikhail~A.~Mishchenko,
        Denis~I.~Bolshakov,
				Alexander~S.~Vasin,
        Valery~V.~Matrosov,
        and~Ilya~V.~Sysoev,~\IEEEmembership{Member,~IEEE}
\thanks{The authors are with the department of theory of oscillations and automatic control, Faculty of Radiophysics of the National Research Lobachevsky State University of Nizhny Novgorod, Ilya~V.~Sysoev is also employed in Saratov State University}
\thanks{Manuscript received February 17, 2021.}}

\markboth{Arxiv.org, February 17, 2021}%
{Mishchenko \MakeLowercase{\textit{et al.}}: Identification of phase-locked loop system from its experimental time series}

\maketitle

\begin{abstract}
Phase-locked loops (PLLs) are now widely used in communication systems and have been a classic system for more than 60 years. Well-known mathematical models of such systems are constructed in a number of approximations, so questions about how they describe the experimental dynamics qualitatively and quantitatively, and how the accuracy of the model description depends on the behavior mode, remain open. One of the most direct approaches to the verification of any model is its reconstruction from the time series obtained in experiment. If it is possible to fit the model to experimental data and the resulting parameter values are close to the expected values (calculated from the first principles), the quantitative correspondence between the model and the physical object is nearly proved. In this paper, for the first time, the equations of the PLL model with a bandpass filter are reconstructed from the experimental signals of the generator in various modes. The reconstruction showed that the model known in the literature generally describes the experimental dynamics in regular and chaotic regimes. The relative error of parameter estimation is between 2\% and 50\% for different regimes and parameters. The reconstructed nonlinear function of phase is not harmonic and highly asymmetric in contrary to the model one.
\end{abstract}


\IEEEpeerreviewmaketitle

\section{Introduction}

Phase-locked loop systems (PLLs) has been widespread in radioengineering for a long time, at lest 60 year \cite{Shakhgildyan_Lyakhovkin_book1972,Best_book,Banerjee_book}, providing large variety \cite{HsiehHun_IEEE1996,Gardner_book3}. They became a classical circuit type and demonstrated many types of behavior \cite{MatrosovShalfeev_book}, including irregular and chaotic \cite{EndoChua1988,ChuChouChang1990,EndoChua1991,Endo1994,ShalfeevMatrosov2002,Piqueira2017}. Significant relevance of PLLs together with their broad range demand permanent analysis and investigation. Most studies now are performed in a form of computer modeling using different SPICE simulators or numerical analysis of mathematical models. The number of experimental works remains relatively small \cite{SatoEndo1995,Mishagin2007,Sarkar2014}. However, one should take into account that mathematical models of PLLs usually have significant limitations being constructed with different approximations and do not mirror the experimentally observed dynamics. Here, we consider the PLL system with bandpass filter \cite{Shalfeev1968} which is interesting for multiple application due to many different regimes it provides \cite{MischenkoAND2012,Matrosov_etal_EPJST2013} and can be considered as a model of both radiophysical circuits and biological neurons, especially after some additional assumptions \cite{Sysoev_etal_TPhL2020}. In this work we aim to reconstruct the proposed in \cite{Shalfeev1968} model from experimental time series of the generator constructed in \cite{Mishchenko_etal_TPhL2017} and detect qualitatively and quantitatively how much a typical PLL mathematical model really describes the experimental setup. 

The idea to reconstruct a system of ODEs from time series of experimental device is not rather new \cite{CremersHuebler_ZeitNaturA1987}. Some first works in this field were very promising \cite{GouesbetLetellier_PRE1994}, proposing many applications, including forecast, coupling analysis, indirect measurement and model verification, that is interesting for us \cite{BezruchkoSmirnovBook}. But then, application of the idea to experimental and even to simulated data were not very successful. As a result, many researcher were disappointed and the progress in the field occurred to be not very fast. The general algorithms like proposed in \cite{GouesbetLetellier_PRE1994} occurred to be inefficient \cite{BezruchkoSmirnov_PRE2000}, model dimension was too large, and amount of experimental data seemed to be not enough to estimate model parameters. Some progress was available when the idea to use a priori information \cite{Anishchenko_etal1998_CSF} was converted into a number of special approaches to particular classes of systems \cite{ProkhorovPonomarenko_PRE2005,SmirnovBezruchko_PRE2009,ShandilyaTimme_NewJPhys2011,Molkov_etal_PRE2011,Han_etal_PRL2015,Brunton_etal_ProcNASUSA2016,Mangan_etal_IEEEMolBiol2016}. Here, we mainly follow the approach for reconstruction of nonlinear oscillators from scalar experimental series proposed in \cite{Sysoev_PhysicaD2018} for van der Pol oscillators with many additional corrections and changes to fit the specifics of the considered system.

\section{Model equations}
Classical PLL system includes a voltage controlled oscillator (VCO), source of external periodic signal --- reference generator (RG), phase detector (PD), and a filter in the control loop. The typical model for such a system is a nonlinear equation for phase difference $\varphi$ between RG and VCO \cite{Shakhgildyan_Lyakhovkin_book1972}:
\begin{equation}
p\varphi+\Omega_HF(\varphi)K(p)=\Delta\omega, \label{eq:PLLgen}
\end{equation}
where $p$ is differentiation operator, $\Omega_H$ is the PLL hold band, $\Delta\omega$ is the difference between RG frequency $\omega_{\mathrm{RG}}$ and uncontrolled (without driving) frequency of VCO $\omega_0$, $F(\varphi)$ --- phase detector characteristics, with maximal valued normalized to one, $K(p)$ --- transfer characteristic of the filter.

If there are frequency dividers in the PLL, the equation (\ref{eq:PLLgen}) should be rewritten as follows:
\begin{equation}\label{eq:PLLdiv}
p\varphi+\frac{\Omega_H}{n}F(\varphi)K(p) = \frac{\omega_{\mathrm{RG}}}{m}-\frac{\omega_0}{n},
z\end{equation}
where $m$ and $n$ are coefficients of oscillation frequency division for RG and VCO respectively.

The studied here experimental generator with bandpass filter constructed in \cite{Mishchenko_etal_TPhL2017} uses logical XOR element as a phase discriminator. The bandpass filter transfer function is described by the formula $K(p)=T_1p(1+T_1p)^{-1}(1+T_2p)^{-1}$, where $T_1=R_1C_1$ and $T_2=R_2C_2$. Then, the dimensionless parameters and renormalized time are provided:
\begin{IEEEeqnarray}{rCl}\label{eq:parameters}
\varepsilon_1 &=& \frac{\Omega_H}{n} R_1 C_1, \nonumber\\
\varepsilon_2 &=& \frac{\Omega_H}{n} R_2 C_2, \\	
\gamma &=& \frac{n}{\Omega_H} (\frac{\omega_{\mathrm{RG}}}{m} - \frac{\omega_0}{n}), \nonumber\\
t_{\mathrm{new}}&=&\frac{\Omega_H}{n}t_{\mathrm{old}}, \nonumber
\end{IEEEeqnarray}
where $R_1$, $C_1$, $R_2$, $C_2$ are loop filter parameters , and $\Omega_H = SE$ is the holding band of PLL, with $S$ being VCO sensitivity, $E$ the maximum possible output amplitude of phase discriminator.

Substituting the transfer function $K(p)$ to the operator equation (\ref{eq:PLLdiv}) and using the introduced dimensionless parameters (\ref{eq:parameters}), one can get the system of equations, describing dynamics of the considered generator (\ref{eq:orig_model}), as it was proposed in \cite{Shalfeev1968}:
\begin{IEEEeqnarray}{rCl}\label{eq:orig_model}
\frac{d\varphi}{dt} &=& y, \nonumber\\
\frac{dy}{dt} &=& z, \\	
\varepsilon_1 \varepsilon_2 \frac{dz}{dt} &=& 
\gamma - (\varepsilon_1 + \varepsilon_2)z -
(1+\varepsilon_1 \cos\varphi)y \nonumber
\end{IEEEeqnarray}
where $\varphi$ is a current phase difference of VCO and RG, $\gamma$ is a relative initial frequency detuning, $\varepsilon_1, \varepsilon_2$ are the filter parameters.

All dimensionless values introduced in (\ref{eq:parameters}) can be calculated when studying the experimental generator, so the parameter ``true'' values of the model (\ref{eq:orig_model}) can be estimated from the first principles.
To write down the mathematical model (\ref{eq:orig_model}) the dimensionless time $t=\frac{\Omega_y}{n}t'$ was introduced. Therefore, the time constant $T_{\mathrm{renorm}}=\Omega_y/n$ have to be estimated to establish compliance between parameters of the model and parameters of the experimental generator.

The phase detector is based on XOR logical element and produces the parasite signal component with the frequency near $(\omega_{ref}/m+\omega_0/n)$. When the equations (\ref{eq:orig_model}) were written down there was suggested that the signal high frequency components including this parasite component are removed by the filter in control loop. However, for low order filters with small inertia parameters like the filter used here, high frequency components are not completely vanished being only partly attenuated. Moreover, the developed generator \cite{Mishchenko_etal_TPhL2017} provides signal shift from $0$ at the filter output in control loop that is also not accounted in the model (\ref{eq:orig_model}).

\section{Reconstruction approach}
\subsection{The previously proposed technique}
The approach to reconstruction of equations (\ref{eq:orig_model}) was proposed in \cite{Sysoeva_etal_PND2020}. In brief, it can be described as follows. The observable (experimentally measured signal) corresponds to the variable $y$ in the equation (\ref{eq:orig_model}). For generality, let us denote $f(\varphi) = 1+\varepsilon_1\cos\varphi$ considering that it is unknown smooth function. Time series of variables $\varphi$ and $z$ can be calculated from the observable $y$ by means of numerical integration and differentiation. Then, let us express the introduce function $f$ from the last equation of system (\ref{eq:orig_model}) and the new ``effective'' parameters $\alpha_0$ and $\alpha_1$:
\begin{IEEEeqnarray}{rCl}
\label{eq:f_phi}	
f(\varphi) &=& \frac{\alpha_0}{y}+ \alpha_1 \frac{z}{y}- \frac{1}{y} \frac{dz}{dt}; \\
\alpha_0 &=& \frac{\gamma}{\varepsilon_1\varepsilon_2}; \nonumber\\
\alpha_1 &=& -\frac{\varepsilon_1+\varepsilon_2}{\varepsilon_1\varepsilon_2}.
\nonumber
\end{IEEEeqnarray}
Time series of the derivative $\frac{dz}{dt}$ can be also obtained using numerical differentiation. So, the four component state vector $(\varphi, y, z, \frac{dz}{dt})$ is obtained from the scalar series of observable.

Then, let us introduce the sorting map $Q(n)$, which make correspondence between the state vector with the number $n$ in the original series the state vector with a number $Q(n)$ in a new series, where all vectors are sorted with an increase of $\varphi$ component. The reverse map can be denoted as $Q^{-1}$, providing $Q^{-1} (Q(n))=n$. Let us consider the vector preceding the vector number $Q(n)$ in the sorted series, which has the number $p_n=Q^{-1}(Q(n)-1)$ in the original one. The function $f$ increment at the segment $[\varphi(p_n);\varphi(n)]$ can be expressed as (\ref{eq:delta}):
\begin{eqnarray}	
\delta_n &=&
- \Delta\dot{z}(n)+\alpha_0 \Delta y^{-1}(n)
+ \alpha_1 \Delta \upsilon(n),\nonumber\\
\Delta\dot{z}(n) &=& \frac{1}{y(n)} \frac{dz}{dt}(n) - \frac{1}{y(p_n)} \frac{dz}{dt}(p_n), \label{eq:delta}\\
\Delta y^{-1}(n) &=& \frac{1}{y(n)} - \frac{1}{y(p_n)},\nonumber\\
\Delta \upsilon(n) &=& \frac{z(n)}{y(n)} -
\frac{z(p_n)}{y(p_n)}\nonumber
\end{eqnarray}
The sum of $\delta_n$ squares is available as a target function, with minimizing which the reconstruction procedure can be reduced to linear least squares routine.
\begin{equation}\label{eq:L}
L(\alpha_0, \alpha_1) = \sum\limits_n \delta_n^2 
\end{equation}

The advantage of the proposed approach to the identification of the model (\ref{eq:orig_model}) in comparison to the direct reconstruction of equation with explicit approximation of the nonlinear function by polynomials \cite{GouesbetLetellier_PRE1994} or by another type of series, as it was mentioned in \cite{Gouesbet_book2003,BezruchkoSmirnovBook}, is that it makes less assumptions and partly reduces the explicit parametrization of $f$. Therefore, it became more robust to inconsistency between the model and real evolution operator that is of great importance for application to real data.

\subsection{The shortcomings of the existing algorithm}
When the described identification technique is applied to experimental data, the already known problems \cite{BezruchkoSmirnovBook} typical for general system identification algorithms also take place.
\begin{enumerate}
	\item The model (\ref{eq:orig_model}) does not completely describe the experimentally measured signal, since there are additional components present in it, well seen in both time series and spectra \cite{Mishchenko_etal_TPhL2017}. Since these components are mostly high frequency they effect the time series of variables $z$ and $\frac{dz}{dt}$ a lot.
	\item The noise level in the experimental setup is higher than the noise appropriate for efficiency of identification technique as considered in \cite{Sysoeva_etal_PND2020}, leading to additional problems in reconstruction due to large corruption of state vector components $z$ and $\frac{dz}{dt}$.
	\item The equations (\ref{eq:orig_model}) were written in \cite{Shalfeev1968} by means of a number of simplifications, including normalization to time constatnt $T_{\mathrm{renorm}}$. If this constant is unknown of accounted with error, the reconstructed values of $\alpha_0$ and $\alpha_1$ would be scaled while the procedure in general would not be completely inoperable.
	\item In the experiment, the linear observation function $\eta$ is measured instead of variable $y$ by itself, and the desired $y$ variable can be expressed through $\eta$ as $y = a \eta + b$.
\end{enumerate}

Due to observable function $\eta$, the state vector components have a form
\begin{IEEEeqnarray}{rCl}
y =& a \eta + b, & \nonumber\\
\psi =& \int \eta dt, & \varphi = a \psi + bt + c, \label{eq:eta_psi_zeta}\\
z =& a \zeta, \nonumber &\\
\frac{dz}{dt} =& a \frac{d\zeta}{dt}, & \nonumber
\end{IEEEeqnarray}
where $\psi$ can be achieved by means of numerical integration and $\zeta$ and $ \frac{d\zeta}{dt}$ are obtained using sequential numerical differentiation of the observable $\eta$.

Additional time scales in the signal not accounted in the model as well as measurement noise can be mostly compensated by means of filtering and smoothing the observable if the sampling rate is quite high. The time constant $T_{\mathrm{renorm}}$ can be estimated base on the setup characteristics as well as the scaling factor $a$. Even if  $T_{\mathrm{renorm}}$ and $a$ would be estimated with some error, this only would lead to some scaling of the resulting parameter values. 
The main problem is an unknown coefficient $b$ since it is both leads to linear trend in $\psi$ due to numerical integration and located in the denominator of the equation (\ref{eq:f_phi}). As a result, we got a nonlinear optimization problem with singularity instead of well defined linear least-squares problem. Additionally, $f$ becomes dependent not only on phase $\phi$, but also on time with unknown coefficient. Taking all this circumstances in mind it is hardly to count for success using the previously developed approach, which was efficient enough for simulated data but occurred to be unsuitable for experimental setup.

\subsection{The technique adapted to experimental data}
The possible way to solve the problems originated from the experimental setup is to switch from the task of original system (\ref{eq:orig_model}) identification to identification of the integrated over time equation (\ref{eq:model_integrated}). Such an approach solves a number of problems described in the previous subsection, as it is explained further. At the same time, it the equation (\ref{eq:model_integrated}) is well identified from series, this definitely means that the original model (\ref{eq:orig_model}) also matches the experimental setup.
\begin{equation}\label{eq:model_integrated}
\varepsilon_1\varepsilon_2 z = \gamma t -(\varepsilon_1+\varepsilon_2)y - \int (1+\varepsilon_1 \cos\varphi)y dt.
\end{equation}
Since the proposed approach is not based on direct representation of the pahe $\varphi$ (only smoothness is necessary) let us denote $f_1(\varphi) = (1+\varepsilon_1 \cos\varphi)$. Given $y = \frac{d\varphi}{dt}$ from the first equation of (\ref{eq:orig_model}), let us rewrite (\ref{eq:model_integrated}) as follows:
\begin{equation}\label{eq:model_integrated1}
\varepsilon_1\varepsilon_2 z = \gamma t -(\varepsilon_1+\varepsilon_2)y 
- \int f_1(\varphi)\frac{d\varphi}{dt}  dt
\end{equation}

It is obvious that the integrand in (\ref{eq:model_integrated1}) is a derivative of a complex function and can be simplified by integrating it:
\begin{equation}
\int f_1(\varphi)\frac{d\varphi}{dt} dt = \int f_1(\varphi) d\varphi = f_2(\varphi), 
\end{equation}
where $f_2(\varphi)$ is a new smooth function of $\varphi$, with integration constant being also included in $f_2$. Then, using notation from (\ref{eq:f_phi}) and introducing $f_3(\varphi) = f_2(\varphi) / (\varepsilon_1\varepsilon_2)$ the equation (\ref{eq:model_integrated1}) can be rewritten in the form (\ref{eq:f3}). 
\begin{equation}\label{eq:f3}
f_3(\varphi) = \alpha_0 t + \alpha_1 y - z.
\end{equation}
The equation (\ref{eq:f3}) has two main advantages over (\ref{eq:f_phi}). First, there is no need any more to numerically estimate the second derivative $\frac{dz}{dt}$ of the observable $y$. Second, there is no unknown parameter in the denominator at the right side of equation. And it is obvious that smoothness properties of $f_3(\varphi)$ are not worse than those of $f$ under the assumptions $\varepsilon_1 \neq 0$ and $\varepsilon_2 \neq 0$.

Let us consider more complex and realistic version of the reconstruction task, when the $y$ variable is measured with a shift $b$, as it was proposed earlier (\ref{eq:eta_psi_zeta}). Then, let us substitute $y = \eta + b$ and $\varphi = \psi + bt$ in (\ref{eq:f3}) and use the Taylor series near 0 (additionally, let us assume that $t=0$ corresponds to the middle of the considered time series: $t \in [-N\Delta t /2; N\Delta t /2]$). In such a case eq.~(\ref{eq:f3}) can be written as:
\begin{equation}\label{eq:f3_Taylor}
\begin{array}{l}
  f_3(\psi) + \left.\frac{df_3(\psi+bt)}{dt}\right|_{t=0}t + \\ 
  \left.\frac{d^2f_3(\psi+bt)}{dt^2}\right|_{t=0}\frac{t^2}{2} + \dots = 
  \alpha_0 t + \alpha_1 \eta + \alpha_1 b - z.  
\end{array}
\end{equation}
If we assume that $b$ is relatively small and limit the series by $K$-th term, we can get the following equation grouping the terms by order of $t$:
\begin{eqnarray}
  f_3(\psi) - \alpha_1 b = \alpha_1 \eta - z + \nonumber\\
  \left(\alpha_0 - \left.\frac{df_3(\psi+bt)}{dt}\right|_{t=0}\right) t -\label{eq:f3_Taylor2}\\
  \sum_{k=2}^{K} \nonumber
  \left.\frac{d^kf_3(\psi+bt)}{dt^k}\right|_{t=0} \frac{t^k}{k!}
\end{eqnarray}
Now, let us denote new functions in (\ref{eq:f3_Taylor2}) which depend on $\psi$:
\begin{eqnarray}
  f_4 (\psi) &=& f_3(\psi) - \alpha_1 b \\
  f_{4,1} (\psi) &=& \alpha_0 - \left.\frac{df_3(\psi+bt)}{dt}\right|_{t=0}\\
  f_{4,k} (\psi) &=& - \frac{1}{k!}\left.\frac{d^kf_3(\psi+bt)}{dt^k}\right|_{t=0},
  k = 2, \dots, K. 
\end{eqnarray}
These new functions does not depend on $t$, only on $\psi$. If one considers Taylor series for these functions and limit with zero term (i.\,e. constant $\psi_0$) such a simplified version of equation (\ref{eq:f3_Taylor2}) can be written as follows:
\begin{eqnarray}
  f_4(\psi) &=& \beta_0 \eta + \sum_{k=1}^K \beta_k t^k - z, \label{eq:f4}\\
  \beta_0 &=& \alpha_1\nonumber,\\
  \beta_1 &=& \alpha_0 - \left.\frac{df_3(\psi_0+bt)}{dt}\right|_{t=0}\nonumber,\\
  \beta_k &=& - \left.\frac{d^kf_3(\psi_0+bt)}{dt^k}\right|_{t=0}\nonumber.
\end{eqnarray}
Using this expression one can construct an approach similar to the proposed in \cite{Sysoeva_etal_PND2020}. Let us sort the state space vectors $(\psi_n, \eta_n, \zeta_n)$ by increase of $\psi$, and let us introduce the difference $\delta_n$:
\begin{eqnarray}
  \delta_n &=& f_4(\psi(n)) - f_4(\psi(p_n)) = \nonumber\\
  &&\beta_0 \Delta\eta(n) + 
  \sum_{k=1}^K \beta_k h_k(n) - \Delta z(n) \label{eq:delta_nov} \\
  \Delta\eta(n) &=& \eta(n) - \eta(p_n) \nonumber\\
  \Delta z(n) &=& z(n) - z(p_n) \nonumber \\
  h_k(n) &=& t_n^k - t_{p_n}^k, \nonumber
\end{eqnarray}
where $n$ and $p_n$ have the same sense as earlier. Using this formalism the target function can be written as (\ref{eq:L}) with the only difference that $\delta$ is taken from (\ref{eq:delta_nov}) rather than from(\ref{eq:delta}). The model identification problem is reduced by these means to the linear least-squares routine, from which $\beta_k$ can be estimated.

\section{Results}
We considered seven different regimes of the experimental generator. Six of them were regular nonlinear oscillations with different number of spikes on a burst, denoted as regimes 1--6 (the number is equal to the number of spikes), and one regime was chaotic. The regime 1 can be considered as a spiking regime, and others including the chaotic one as bursting regimes. To provide a clear matching between the reconstructed regime and model regimes described earlier in \cite{Matrosov_etal_EPJST2013} we kept literal abbreviations from ``(b)'' to ``(f)'' used in fig.~1 of \cite{Matrosov_etal_EPJST2013}. Two additional regimes of experimental generator which do not have matching model regimes reported previously have no literal abbreviations. Regular quasi-linear regime ``(a)'' is not considered, since it caries too few information for correct identification as it was shown even from the simulated data in \cite{Sysoeva_etal_PND2020}.
\begin{figure*}
	\includegraphics[width=0.49\linewidth]{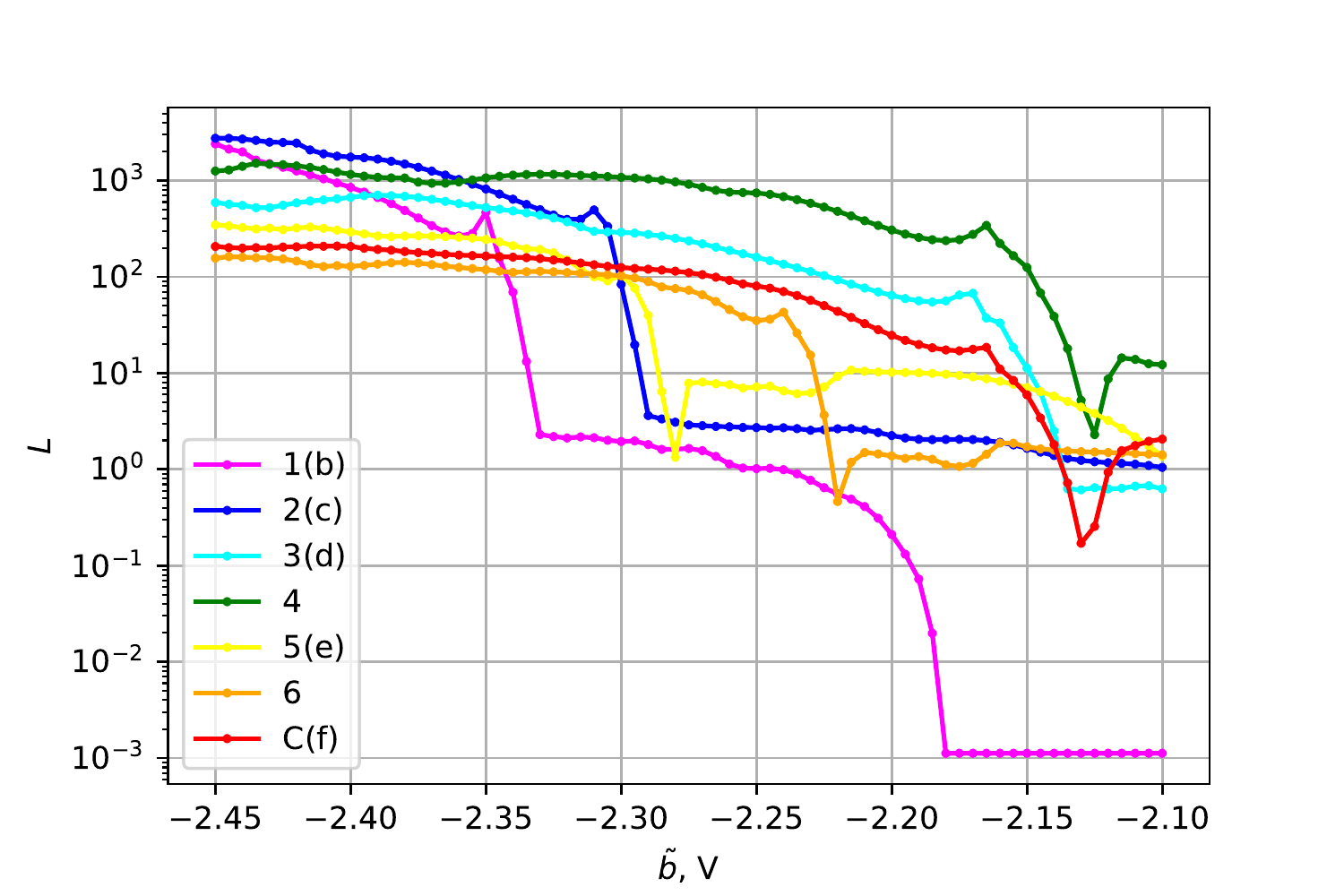}
	\includegraphics[width=0.49\linewidth]{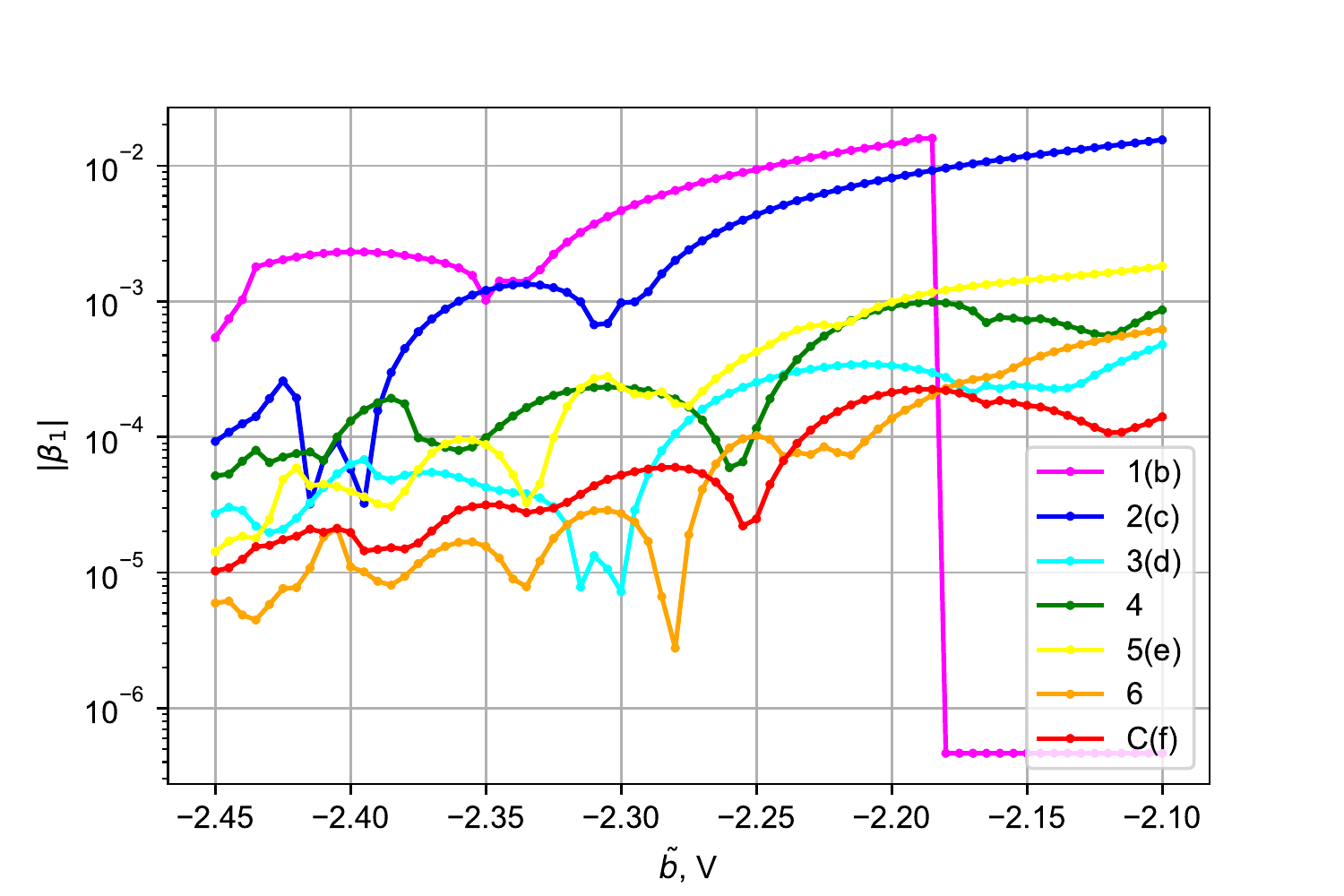}\\
	\centerline{(a)\hspace{0.4\linewidth}(b)}
	\caption{Dependency of target function value (\ref{eq:L}) (part a) and absolute value of $\beta_1$ coefficient (part b) from trial shift $\tilde{b}$. 
Different colors correspond to different dynamical regimes: regular oscillations with 1--6 spikes in a burst (denoted using numbers) and chaotic regime (denoted as ``C''). A small letter in brackets is match the considered regime with model regime described in detail in \cite{Matrosov_etal_EPJST2013}, see fig.~1 of that paper.}\label{fig:Lbeta_b}
\end{figure*}

To detect the measurement shift $b$ the identification was performed for different trial shifts $\tilde{b}$. The range of trial $\tilde{b}$ to be studied can be estimated by detailed investigation of regimes in the model (\ref{eq:orig_model}). The dependencies of the target function $L$ and coefficient  $\beta_1$ on $\tilde{b}$ for different dynamical regimes were plotted in fig.~\ref{fig:Lbeta_b} with various colors. In the legend the numbers 1--6 mean the number of spikes on a burst for regular regimes and the letter ``C'' indicates the chaotic regime. A small letter in brackets is to provide coincidence with the model regimes depicted in fig.~1 of the paper \cite{Matrosov_etal_EPJST2013}. The dependencies $L(\tilde{b})$ have the sharp slope, right to which the $L$ values fall $\sim 10^2$ times. The additional analysis showed the monotonous increase of $\psi$ estimated using $\tilde{b}$ values right to the slope. Due to such a monotonous behavior, the sorting map $Q$ looses any sense and the whole identification algorithm occurs to be inefficient since the approximating terms at $\beta_i$ in (\ref{eq:f4}) become small by themselves rather than in a linear combination; i.\,e. any values of $\beta_i$ give similar results. Such a behavior of $\psi$ is far from the natural (we assume that for $\tilde{b}$ matching the actual $b$ value $\psi=\phi$ and therefore one can transfer results of previous model analysis here). Therefore, the values of $\tilde{b}$ at the slope and right to it are not considered.

To estimate $b$ we propose using the most right minimum of the dependency $|\beta_1|(\tilde{b})$ (see fig.~\ref{fig:Lbeta_b}b) which is still left to the slope. The following idea underlies this technique. First, the $\gamma$ values are usually relatively small, and therefore $|\beta_1|$ should be also relatively small since the derivative $\frac{df_3}{dt}$ make the main impact into it, see eq.~\ref{eq:f4}. This derivative should become zero if the actual value of shift $b$ is considered. Second, the left minima of $|\beta_1|(\tilde{b})$ dependency correspond to high values of the target function.

\begin{figure}
	\includegraphics[width=0.99\linewidth]{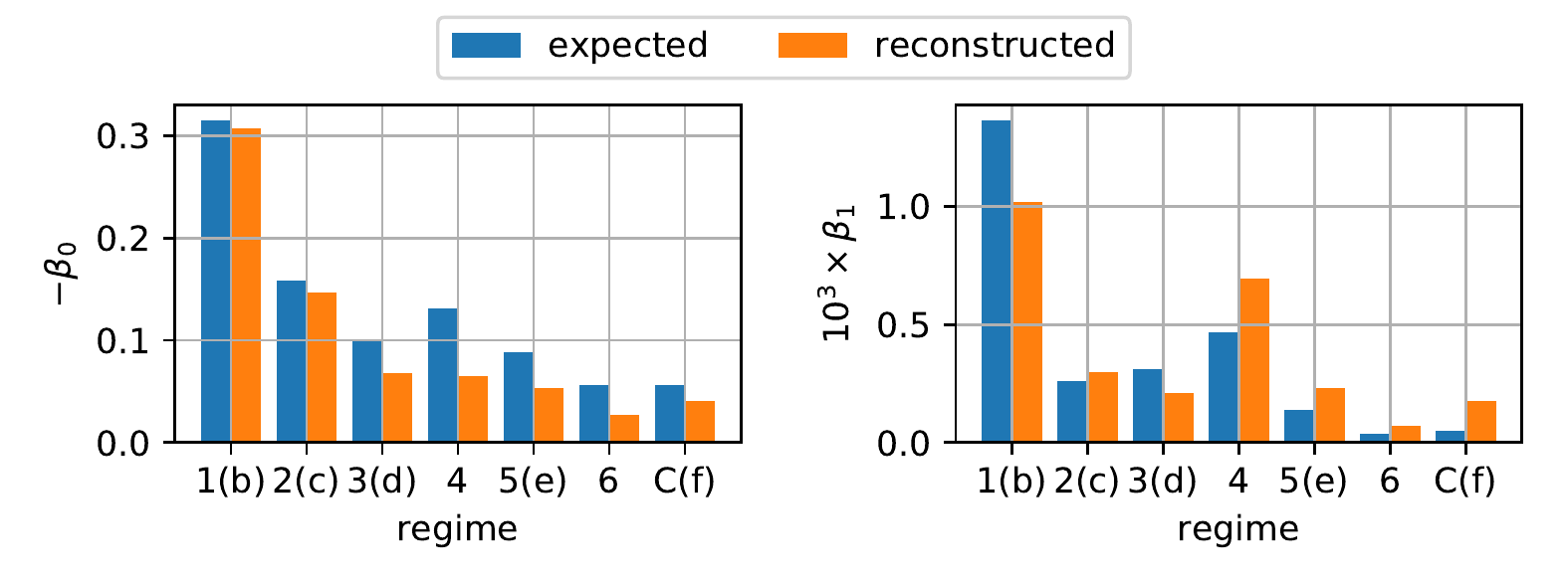}
	\caption{Estimated parameter $\beta_0$ and $\beta_1$ values (orange) and their theoretically expected values (blue), calculated based on nominal values of elements used for setup construction.}\label{fig:param}
\end{figure}
The reconstructed values of parameters $\beta_0$ and $\beta_1$ were plotted on fig.~\ref{fig:param} in comparison with their theoretically expected values. The results of $\beta_0$ reconstruction show relatively small error (from 2\% to 52\%) in different regimes. The largest error is for the regime 6, while $\beta_0$ reconstructed for the chaotic regime, which is the most complex one, has an error similar to the regime with 3 spikes. It is also interesting that there is now strict dependency between regime complexity and error in reconstruction. Also, one have to notice that there could be a mismatch (up to 5\%) in theoretically expected values of $\beta_0$ and $\beta_1$ calculated from nominal values of components and actual effective value.

The relative mismatches between $\beta_1$ reconstructed from series and calculated theoretically is mostly much larger than for $\beta_0$. There are two main sources of this mismatch. First, value of $\gamma$ is usually very small and close to zero. Therefore, it is hard to estimate $\beta_1$ which is proportional to $\gamma$ precisely. Second, we used the dependency $\beta_1(\tilde{b})$ as a way to estimate the unknown constant shift assuming that the minimum corresponds to it. But this technique is somewhat biased since we limited an approach by the zero term in the Taylor series for the function $\left.\frac{df_3(\psi_0+bt)}{dt}\right|_{t=0}$.

\begin{figure}
	\includegraphics[width=0.99\linewidth]{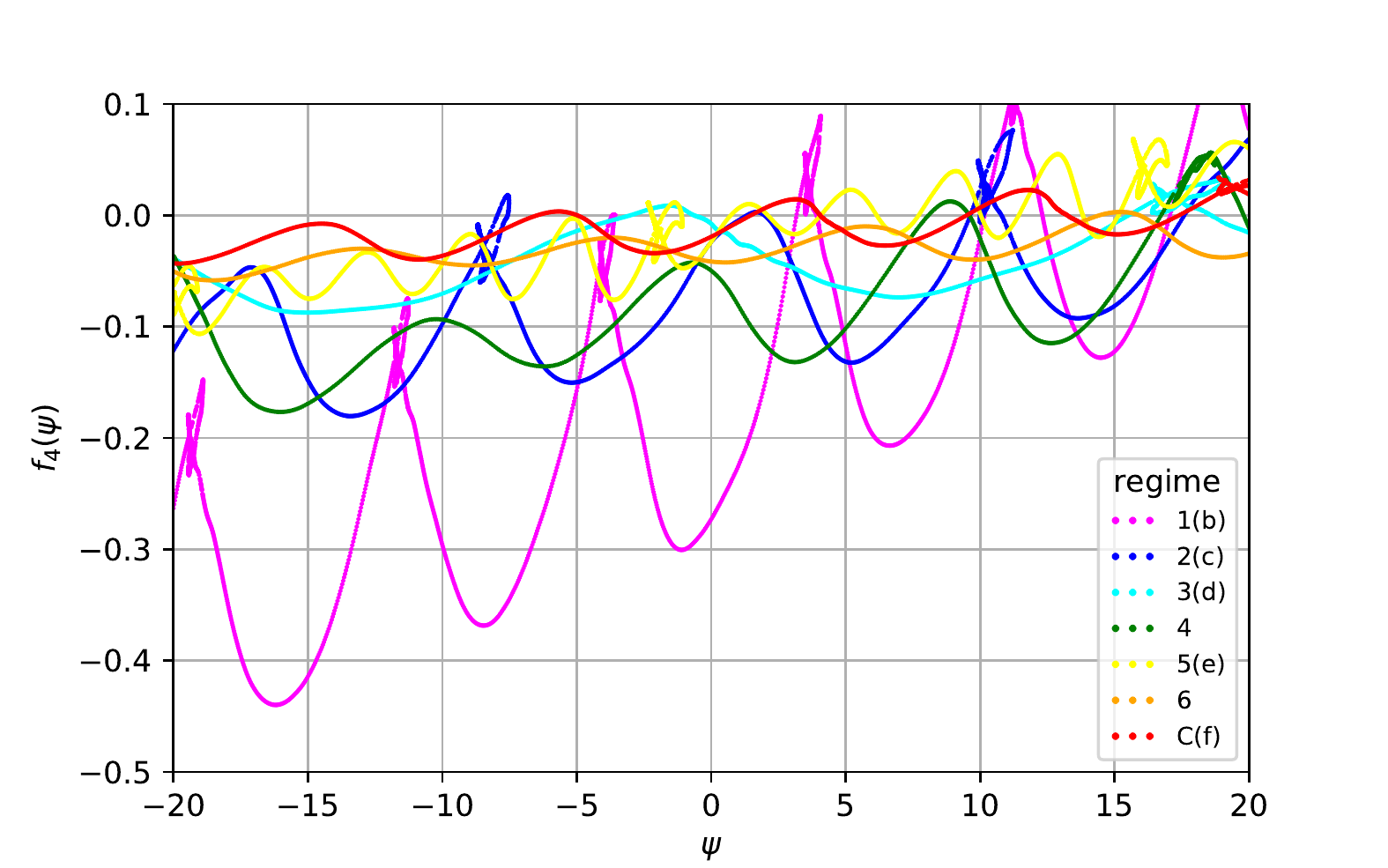}
	\caption{Reconstructed in different regimes nonlinear functions $f_4(\psi)$ (notation is the same as in the fig.~\ref{fig:Lbeta_b}).}\label{fig:f4_psi}
\end{figure}

Results of nonlinear function  $f_4(\psi)$ reconstruction are shown in the fig.~\ref{fig:f4_psi}. In general the reconstructed function is more or less close to the sine with a linear trend as it follows from the models  (\ref{eq:orig_model},\ref{eq:model_integrated1}). However, there are noticeable differences:
\begin{enumerate}
  \item there are significant ambiguities at many maxima; the general rule taken empirically is that there is ambiguity for each $n$-th maximum for regular regimes with $n$ spikes on a burst;
  \item amplitude of a function is changing for all regimes except 1(b) (simple spiking): it is the smallest after the ambiguous maximum and then grows;
  \item the form is significantly different from the harmonic one even for the simplest case --- spiking regime 1(b).
\end{enumerate}
Based on this analysis one have to admit that assuming $f(\varphi)$ in the model (\ref{eq:orig_model}) to be a harmonic function of a kind $f(\varphi) = 1+\varepsilon_1\cos\varphi$ or even of more general kind $f(\varphi) = c_0+c_1\cos(\varphi+\varphi_0)$ seems to be unappropriated. Therefore, we have to state that the equations (\ref{eq:orig_model}) do not completely match the experimental setup and the idea not to rely on the explicit function form taken from model for system identification was right. If we would ground the method on the explicit approximation of the function from (\ref{eq:orig_model}), the errors in estimated values of parameters would be much higher.

\section{Conclusion and discussion}
I this work two significant novel results were achieved. First, the quantitative correspondence between PLL model proposed in \cite{Shalfeev1968} and its hardware realization (experimental setup) constructed in \cite{Mishchenko_etal_TPhL2017} was established. This is a first time when the PLL model was verified by its direct identification (reconstruction) from experimental data. Second, the more general result is that the idea of model verification by means of its direct reconstruction from experimental data proposed in the number of works \cite{Gouesbet_book2003,BezruchkoSmirnovBook} was tested and shown to be fruitful for real physical devices. 

The results of identification demonstrated that the model (\ref{eq:orig_model}) fits the experimental series in some regimes better than in others. Also, one has to take into account both linear transform of the observable and additional high frequency component we filtered out before identification technique application.

The success of the study became possible due to complete take into account all known a priori information about both the model and the setup. The previously proposed and tested for simulated data in different dynamical regimes approach \cite{Sysoeva_etal_PND2020} occurred to be insufficient and not directly applicable to experimental data. To solve this problem, the significant modification of the proposed technique was developed. This technique includes analytical integration of model equations and identification of integrated equation instead of the original one. Such an approach was proposed for the first time. For the particular system, it allows to avoid additional numerical differentiation, since time series of the second derivative was not necessary. Also, it partly solved a problem of linear measurement transform (observation function). 

The solved problem was very complex, since we reconstructed equations of 3rd order ODE system from experimental series using scalar series of linearly scaled variable with unknown constant shift. Previously, hidden variable approaches were considered to solve problems of such a complexity \cite{BBBB_PRA1992}, but they were tested and applied mostly to simulated data. And even for such data researchers faced a lot of problems \cite{Bezruchko_etal_CSF2006}. 

Since the proposed PLL systems was shown to demonstrate behavior similar to real neurons \cite{MischenkoAND2012,Matrosov_etal_EPJST2013,Mishchenko_etal_TPhL2017},
we hope that the proposed method may become a first step to identification of neuron models from their experimental extra or intracellular recordings. This can be fruitful for different purposes, including indirect measurement of parameters of different cells, which are now assumed only averagely \cite{Ching_etal_PNAS2010}, classification of cell types and distinguishing between normal and pathological cells in case of most neurological diseases like epilepsy, Parkinson and Alzheimer. We understand that there is a significant distance between identification of a model from experimental series of radioengineering setup, even if the model neglects some significant features of the device, and identification of biological systems. So, there is a lot to do steel. However, we hope that the step done in frames of the current work is valuable and fruitful for further success.

\appendices
\section{Parameter values}
All values of experimental setup parameters are presented in the table~\ref{tab:1}. The detailed results of parameter reconstruction are presented in table~\ref{tab:2}.
\begin{table}[h!]
	\setlength{\tabcolsep}{1.0mm}
	\caption{Table 1. Parameters of experimental setup}
	\centering
	\begin{tabular}{|p{0.7cm}|p{0.6cm}|l|p{0.5cm}|l|p{0.7cm}|p{0.5cm}|p{0.5cm}|l|l|l|l|} \hline
		Re\-gime & $\omega_{RG}$, MHz & m & $\omega_0$, MHz & n & $\Omega_H$, Mrad/s & $R_1$, Ohm & $R_2$, Ohm & $\gamma$ & $\varepsilon_1$ & $\varepsilon_2$ \\ \hline
		1(b) & 16 & 17000 & 5 & 5000 & 29.8 & 2000 & 4000 & 0.062 & 4.77 & 9.53 \\ \hline
		2(c) & 16 & 17000 & 10 & 10000 & 83.9 & 3000 & 5000 & 0.044 & 10.1 & 16.8 \\ \hline
		3(d) & 16 & 8000 & 8 & 3500 & 46.9 & 3000 & 5000 & 0.134 & 16.1 & 26.8 \\ \hline
		4 & 16 & 8000 & 8 & 3500 & 46.9 & 2000 & 5000 & 0.134 & 10.7 & 26.8 \\ \hline	
		5(e) & 16 & 18100 & 10 & 10000 & 100 & 6500 & 5000 & 0.0726 & 26.1 & 20.1 \\ \hline
		6 & 16 & 7500 & 8 & 3500 & 70.2 & 4000 & 5000 & 0.0477 & 32.1 & 40.1 \\ \hline
		C(f) & 16 & 7700 & 8 & 3500 & 70.2 & 4000 & 5000 & 0.0651 & 32.1 & 40.1 \\ \hline
	\end{tabular}
	\label{tab:1}
\end{table}

\begin{table}
	\caption{Table 2. Expected values of $\beta_0$ and $\beta_1$, calculated based on nominal values of elements used for experimental setup construction, and estimated values}
	\setlength{\tabcolsep}{1.0mm}
	\centering
	\begin{tabular}{|p{0.7cm}|l|p{0.9cm}|p{0.9cm}|p{1.1cm}|l|l|l|} \hline
		Re- & \multirow{2}{*}{$T_{\mathrm{renorm}}$} & \multirow{2}{*}{$a$} & \multirow{2}{*}{$b$} & \multicolumn{2}{|c|}{Expected} & \multicolumn{2}{|c|}{Estimated} \\ 
		 \cline{5-8}gime&&&&$-\beta_0$ &  $\beta_1 \cdot 10^3$ &  $-\beta_0$ & $\beta_1 \cdot 10^3$ \\ \hline
		1(b) & 5960 & 0.6197 & $-2.35$ & $0.31457$ & $1.363$ & $0.30754$ & $1.020$ \\ \hline
		2(c) & 8390 & 0.4131 & $-2.31$ & $0.15853$ & $0.259$ & $0.14715$ & $0.298$ \\ \hline
		3(d) & 13400 & 0.6197 & $-2.17$ & $0.09943$ & $0.311$ & $0.06819$ & $0.210$ \\ \hline
		4 & 13400 & 0.6197 & $-2.165$ & $0.13077$ & $0.467$ & $0.06479$ & $0.697$ \\ \hline	
		5(e) & 10000 & 0.3443 & $-2.3$ & $0.08807$ & $0.138$ & $0.05293$ & $0.232$ \\ \hline
		6 & 20057 & 0.41 & $-2.24$ & $0.05609$ & $0.037$ & $0.02689$ & $0.072$ \\ \hline
		C(f) & 20057 & 0.41 & $-2.165$ & $0.05609$ & $0.051$ & $0.04024$ & $0.175$ \\ \hline
	\end{tabular}
	\label{tab:2}
\end{table}

\section*{Acknowledgment}
Mikhail~A.~Mishchenko, Denis~I.~Bolshakov, Alexander~S.~Vasin and~Ilya~V.~Sysoev thank Lobachevsky State University of Nizhny Novgorod for special support of this work within 5-100 academic excellence program.

\ifCLASSOPTIONcaptionsoff
  \newpage
\fi

\bibliographystyle{IEEEtran}
\bibliography{Shalfeev_neuron_reconstruction}

\begin{thebibliography}{10}
\providecommand{\url}[1]{#1}
\csname url@samestyle\endcsname
\providecommand{\newblock}{\relax}
\providecommand{\bibinfo}[2]{#2}
\providecommand{\BIBentrySTDinterwordspacing}{\spaceskip=0pt\relax}
\providecommand{\BIBentryALTinterwordstretchfactor}{4}
\providecommand{\BIBentryALTinterwordspacing}{\spaceskip=\fontdimen2\font plus
\BIBentryALTinterwordstretchfactor\fontdimen3\font minus
  \fontdimen4\font\relax}
\providecommand{\BIBforeignlanguage}[2]{{%
\expandafter\ifx\csname l@#1\endcsname\relax
\typeout{** WARNING: IEEEtran.bst: No hyphenation pattern has been}%
\typeout{** loaded for the language `#1'. Using the pattern for}%
\typeout{** the default language instead.}%
\else
\language=\csname l@#1\endcsname
\fi
#2}}
\providecommand{\BIBdecl}{\relax}
\BIBdecl

\bibitem{Shakhgildyan_Lyakhovkin_book1972}
V.~V. Shakhgildyan and A.~Lyakhovich~A.,
  \emph{\BIBforeignlanguage{Russian}{Phase-locked loop systems}}.\hskip 1em
  plus 0.5em minus 0.4em\relax Moscow: Svyaz, 1972.

\bibitem{Best_book}
R.~E. Best, \emph{Phase-Locked Loops: Design, Simulation, and Applications},
  5th~ed.\hskip 1em plus 0.5em minus 0.4em\relax New York: McGraw-Hill, 2003.

\bibitem{Banerjee_book}
D.~Banerjee, \emph{PLL performance, simulation and design}, 4th~ed.\hskip 1em
  plus 0.5em minus 0.4em\relax Indianapolis: Dog Ear Publishing, 2006.

\bibitem{HsiehHun_IEEE1996}
G.-C. Hsieh and J.~C. Hung, ``Phase-locked loop techniques. a survey,''
  \emph{IEEE Transactions on Industrial Electronics}, vol.~43, no.~6, pp.
  609--615, 1996.

\bibitem{Gardner_book3}
F.~M. Gardner, \emph{Phaselock techniques}, 3rd~ed.\hskip 1em plus 0.5em minus
  0.4em\relax John Wiley \& Sons, 2005.

\bibitem{MatrosovShalfeev_book}
V.~V. Matrosov and V.~D. Shalfeev, \emph{Coupled Phase-locked Loops: Stability,
  Synchronization, Chaos and Communication with Chaos}.\hskip 1em plus 0.5em
  minus 0.4em\relax Singapore: World Scientific, 2018.

\bibitem{EndoChua1988}
T.~Endo and C.~L., ``Chaos from phase-locked loops,'' \emph{IEEE Transactions
  on Circuits Systems}, vol.~35, no.~8, p. 987–1003, 1988.

\bibitem{ChuChouChang1990}
Y.~Chu, J.~Chou, and S.~Chang, ``Chaos from third-order phase-locked loops with
  a slowly varying parameter,'' \emph{IEEE Transactions on Circuits Systems},
  vol.~31, no.~9, p. 1104–1115, 1990.

\bibitem{EndoChua1991}
T.~Endo and L.~O. Chua, ``Synchronizing chaos from electronic phase-locked
  loops,'' \emph{International Journal of Bifurcation and Chaos}, vol.~1,
  no.~03, pp. 701--710, 1991.

\bibitem{Endo1994}
E.~T., ``A review of chaos and nonlinear dynamics in phase-locked loops,''
  \emph{J. Franklin Inst.}, vol.~32, no.~95, p. 859–902, 1994.

\bibitem{ShalfeevMatrosov2002}
V.~Shalfeev and V.~Matrosov, ``Dynamical chaos in phase-locked loops,'' in
  \emph{Chaos in Circuits and Systems}, Singapore, 2002, pp. 111--130.

\bibitem{Piqueira2017}
J.~R.~C. Piqueira, ``Hopf bifurcation and chaos in a third-order phase-locked
  loop,'' \emph{Commun. Nonlinear Sci. Numer. Simul.}, vol.~42, p. 178–186,
  2017.

\bibitem{SatoEndo1995}
A.~Sato and T.~Endo, ``Experiments of secure communications via chaotic
  synchronization of phase-locked loops,'' \emph{IEICE transactions on
  fundamentals of electronics, communications and computer sciences}, vol.~78,
  no.~10, pp. 1286--1290, 1995.

\bibitem{Mishagin2007}
K.~G. Mishagin, V.~V. Matrosov, V.~D. Shalfeev, and V.~V. Shokhnin,
  ``Generation of chaotic oscillations in the experimental scheme of two
  cascade-coupled phase systems,'' \emph{Journal of Communications Technology
  and Electronics}, vol.~52, no.~10, pp. 1146--1152, 2007.

\bibitem{Sarkar2014}
B.~C. Sarkar and S.~Chakraborty, ``Self-oscillations of a third order pll in
  periodic and chaotic mode and its tracking in a slave pll,'' \emph{Commun.
  Nonlinear Sci. Numer. Simul.}, vol.~19, no.~3, p. 738–749, 2014.

\bibitem{Shalfeev1968}
V.~D. Shalfeev, ``Investigation of the dynamics of a system of automatic phase
  control of frequency with a coupling capacitor in the control loop,''
  \emph{Radiophys Quantum Electron}, vol.~11, no.~3, p. 221–226, 1968.

\bibitem{MischenkoAND2012}
M.~A. Mishchenko, V.~D. Shalfeev, and V.~V. Matrosov, ``Neuron-like dynamics in
  phase-locked loop,'' \emph{Izvestiya VUZ. Applied Nonlinear Dynamics},
  vol.~20, no.~4, pp. 122--130, 2012.

\bibitem{Matrosov_etal_EPJST2013}
V.~V. Matrosov, M.~A. Mishchenko, and V.~D. Shalfeev, ``Neuron-like dynamics of
  a phase-locked loop,'' \emph{The European Physical Journal Special Topics},
  vol. 222, p. 2399–2405, 2013.

\bibitem{Sysoev_etal_TPhL2020}
I.~V. Sysoev, M.~V. Sysoeva, V.~I. Ponomarenko, and M.~D. Prokhorov,
  ``Neuron-like dynamics in a phase-locked loop system with delayed feedback,''
  \emph{Tech. Phys. Lett.}, vol.~46, pp. 710--712, 2020.

\bibitem{Mishchenko_etal_TPhL2017}
M.~A. Mishchenko, D.~I. Bolshakov, and V.~Matrosov, ``Instrumental
  implementation of a neuronlike generator with spiking and bursting dynamics
  based on a phase-locked loop,'' \emph{Tech. Phys. Lett.}, vol.~43, p.
  596–599, 2017.

\bibitem{CremersHuebler_ZeitNaturA1987}
J.~Cremers and A.~Hübler, ``Construction of differential equations from
  experimental data,'' \emph{Zeitschrift für Naturforschung - Section A
  Journal of Physical Sciences}, vol.~42, no.~8, pp. 797--802, 1987.

\bibitem{GouesbetLetellier_PRE1994}
G.~Gouesbet and C.~Letellier, ``Global vector-field reconstruction by using a
  multivariate polynomial l2 approximation on nets,'' \emph{Physical Review E},
  vol.~49, no.~6, pp. 4955--4972, 1994.

\bibitem{BezruchkoSmirnovBook}
B.~P. Bezruchko and D.~A. Smirnov, \emph{Extracting Knowledge From Time Series:
  An Introduction to Nonlinear Empirical Modeling}, ser. Springer Series in
  Synergetics.\hskip 1em plus 0.5em minus 0.4em\relax New York: Springer, 2010.

\bibitem{BezruchkoSmirnov_PRE2000}
B.~P. Bezruchko and D.~Smirnov, ``Constructing nonautonomous differential
  equations from experimental time series,'' \emph{Phys. Rev. E}, vol.~63,
  no.~1, p. 016207, 2000.

\bibitem{Anishchenko_etal1998_CSF}
V.~S. Anishchenko, A.~Pavlov, and N.~Janson, ``Global reconstruction in the
  presence of a priori information,'' \emph{Chaos, Solitons \& Fractals},
  vol.~9, no.~8, pp. 1267--1278, 1998.

\bibitem{ProkhorovPonomarenko_PRE2005}
M.~Prokhorov and V.~Ponomarenko, ``Estimation of coupling between time-delay
  systems from time series,'' \emph{Phys. Rev. E}, vol.~72, p. 016210, 2005.

\bibitem{SmirnovBezruchko_PRE2009}
D.~A. Smirnov and B.~P. Bezruchko, ``Detection of coupling in ensembles of
  stochastic oscillators,'' \emph{Phys. Rev. E}, vol.~79, p. 046204, 2009.

\bibitem{ShandilyaTimme_NewJPhys2011}
S.~G. Shandilya and M.~Timme, ``Inferring network topology from complex
  dynamics,'' \emph{New Journal of Physics}, vol.~13, no.~1, p. 013004, 2011.

\bibitem{Molkov_etal_PRE2011}
Y.~I. Molkov, D.~N. Mukhin, E.~M. Loskutov, R.~I. Timushev, and A.~M. Feigin,
  ``Prognosis of qualitative system behavior by noisy, nonstationary, chaotic
  time series,'' \emph{Phys. Rev. E}, vol.~84, p. 036215, 2011.

\bibitem{Han_etal_PRL2015}
X.~Han, Z.~Shen, W.-X. Wang, and Z.~Di, ``Robust reconstruction of complex
  networks from sparse data,'' \emph{Phys. Rev. Lett.}, vol. 114, p. 28701,
  2015.

\bibitem{Brunton_etal_ProcNASUSA2016}
S.~Brunton, J.~Proctor, and J.~Kutz, ``Discovering governing equations from
  data by sparse identification of nonlinear dynamical systems,'' \emph{Proc.
  Natl. Acad. Sci. U. S. A.}, vol. 113, p. 3932–7, 2016.

\bibitem{Mangan_etal_IEEEMolBiol2016}
N.~Mangan, S.~Brunton, J.~Proctor, and J.~Kutz, ``Inferring biological networks
  by sparse identification of nonlinear dynamics,'' \emph{IEEE Trans. Mol.
  Biol. Multi-Scale Commun.}, vol.~2, p. 52–63, 2016.

\bibitem{Sysoev_PhysicaD2018}
I.~V. Sysoev, ``Reconstruction of ensembles of generalized van der pol
  oscillators from vector time series,'' \emph{Physica D: Nonlinear Phenomena},
  vol. 384--385, pp. 1--11, 2018.

\bibitem{Sysoeva_etal_PND2020}
M.~V. Sysoeva, I.~V. Sysoev, V.~I. Ponomarenko, and M.~D. Prokhorov,
  ``Reconstructing the neuron-like oscillator equations modeled by a
  phase-locked system with delay from scalar time series,'' \emph{Izvestiya
  VUZ. Applied Nonlinear Dynamics}, vol.~28, no.~4, pp. 397--413, 2020.

\bibitem{Gouesbet_book2003}
G.~Gouesbet, G.~Meunier-Guttin-Cluzel, and O.~Menard, \emph{Chaos and its
  Reconstruction}.\hskip 1em plus 0.5em minus 0.4em\relax New York: Nova
  Science Publishers, 2003.

\bibitem{BBBB_PRA1992}
E.~Baake, M.~Baake, H.~G. Bock, and K.~M. Briggs, ``Fitting ordinary
  differential equations to chaotic data,'' \emph{Phys. Rev. A}, vol.~45,
  no.~8, pp. 5524--5529, 1992.

\bibitem{Bezruchko_etal_CSF2006}
B.~P. Bezruchko, D.~A. Smirnov, and I.~V. Sysoev, ``Identification of chaotic
  systems with hidden variables (modified bock's algorithm),'' \emph{Chaos,
  Solitons \& Fractals}, vol.~29, no.~1, pp. 82--90, 2006.

\bibitem{Ching_etal_PNAS2010}
S.~Ching, A.~Cimenser, P.~L. Purdon, E.~N. Brown, and N.~J. Kopell,
  ``Thalamocortical model for a propofol-induced $\alpha$-rhythm associated
  with loss of consciousness,'' \emph{Proceedings of the National Academy of
  Sciences}, vol. 107, no.~52, pp. 22\,665--22\,670, 2010.

\end{thebibliography}

\end{document}